\documentclass[aps,prb,twocolumn,groupedaddress]{revtex4-1}

\usepackage[utf8]{inputenc}
\usepackage{graphicx}
\usepackage{dcolumn}
\usepackage{bm}

\begin{document}


\title{Scattering mechanisms of highest-mobility InAs/Al$_{x}$Ga$_{1-x}$Sb quantum wells}


\author{T. Tschirky}
\email[]{tschthom@phys.ethz.ch}
\affiliation{Laboratory for Solid State Physics, ETH Zürich, 8093 Zürich, Switzerland}

\author{S. Mueller}
\affiliation{Laboratory for Solid State Physics, ETH Zürich, 8093 Zürich, Switzerland}

\author{Ch. A. Lehner}
\affiliation{Laboratory for Solid State Physics, ETH Zürich, 8093 Zürich, Switzerland}

\author{S. Fält}
\affiliation{Laboratory for Solid State Physics, ETH Zürich, 8093 Zürich, Switzerland}

\author{T. Ihn}
\affiliation{Laboratory for Solid State Physics, ETH Zürich, 8093 Zürich, Switzerland}

\author{K. Ensslin}
\affiliation{Laboratory for Solid State Physics, ETH Zürich, 8093 Zürich, Switzerland}

\author{W. Wegscheider}
\affiliation{Laboratory for Solid State Physics, ETH Zürich, 8093 Zürich, Switzerland}

\date{\today}
 
\begin{abstract}
We study molecular beam epitaxially grown, undoped Al$_{x}$Ga$_{1-x}$Sb/InAs/AlSb quantum wells with different buffer and barrier designs and varying quantum well width. The highest mobilities were achieved with Al$_{0.33}$Ga$_{0.67}$Sb buffers and lower barriers and a quantum well width of 24~nm. These quasi-single-interface InAs/AlSb quantum well devices reached a gate-tuned mobility of 2.4~$\times~10^{6}$~cm$^2$/Vs at a density of 1~$\times~10^{12}$~cm$^{-2}$ and 1.3~K. In Hall bar devices boundary scattering is found to strongly influence the mobility determination in this mobility regime. Ionized background impurity scattering at low electron densities, device boundary scattering at intermediate electron densities, and intersubband scattering at high electron densities were identified as the most likely dominant scattering processes. Ringlike structures in the Landau fan can be explained using a single-particle model of crossing Landau levels.
\end{abstract}

\pacs{73.21.Fg}

\maketitle

\section{Introduction}

Heterostructures containing InAs have been studied due to their potential applications in high-speed, low power-electronics, such as in heterostructure field effect transistors (HFETs)\cite{Yeh2013} and THz imaging and sensing \cite{Brandstetter2016}. The large band offset between InAs and the AlSb barriers results in excellent carrier confinement and enhanced radiation tolerance\cite{Weaver2005}. Its narrow band gap and strong spin-orbit coupling makes the system ideal for spintronic devices research \cite{Datta1990,Zutic2004,Awschalom2007}. In recent years research on InAs quantum wells has gained significance due to their similarity to InAs/GaSb composite quantum wells for topological insulators \cite{Liu2008} and due to new prospects for realizing a topological superconducting phase supporting Majorana fermions when combined with $s$-wave superconductors \cite{Sau2010,Alicea2010,Lutchyn2010}.

The carrier mobility of InAs quantum wells\cite{Nguyen1993,Thomas1997,Shojaei2015,Shojaei2016a} has for a long time been confined to regions below $1~\times~10^6$~cm$^2$/Vs, whereas GaAs/AlGaAs heterostructures can reach mobility values above $3~\times~10^7$~cm$^2$/Vs\cite{Riedi2016,Gardner2016,Schlom2010,Umansky2009} despite their higher effective mass. This implies that there is ample room for improving the growth techniques and structure designs of InAs quantum wells.

The recent availability of high-quality, almost lattice-matched GaSb substrates has led to a steep increase of growth quality and the subsequent carrier mobilities\cite{Shojaei2015}. Shojaei \textit{et al.}\cite{Shojaei2016a} recently showed that the mobilities of InAs quantum wells with Al$_{0.8}$Ga$_{0.2}$Sb barriers and well widths of up to 15 nm scaled in accordance with Coulombic scattering from charged defects at low electron density. For high densities the gate-tuned mobility of those structures was limited to 0.75~$\times~10^{6}$~cm$^2$/Vs by interface roughness scattering and alloy disorder scattering. In this work we show that by further optimizing structural design parameters, the mobility can be drastically increased. The highest low-temperature mobility achieved was 2.402~$\times~10^{6}$~cm$^2$/Vs at a carrier density of 10.19~$\times~10^{11}$~cm$^{-2}$, which is more than double the mobility of any other published InAs quantum well. We present temperature and density dependent magnetotransport measurements on these samples to identify the remaining limiting scattering mechanisms in this mobility regime. 

\section{Samples and ungated measurements}

All studied samples were grown in a modified Veeco Gen II molecular beam epitaxy (MBE) system. The protective oxide was entirely thermally desorbed from the GaSb substrates inside the MBE at 540~$^\circ$C. All temperatures were measured using a BandIT system with blackbody radiation fitting and the growth rates were calibrated using reflection high-energy electron diffraction (RHEED). 

 \begin{figure}
 \includegraphics[width=0.49\textwidth]{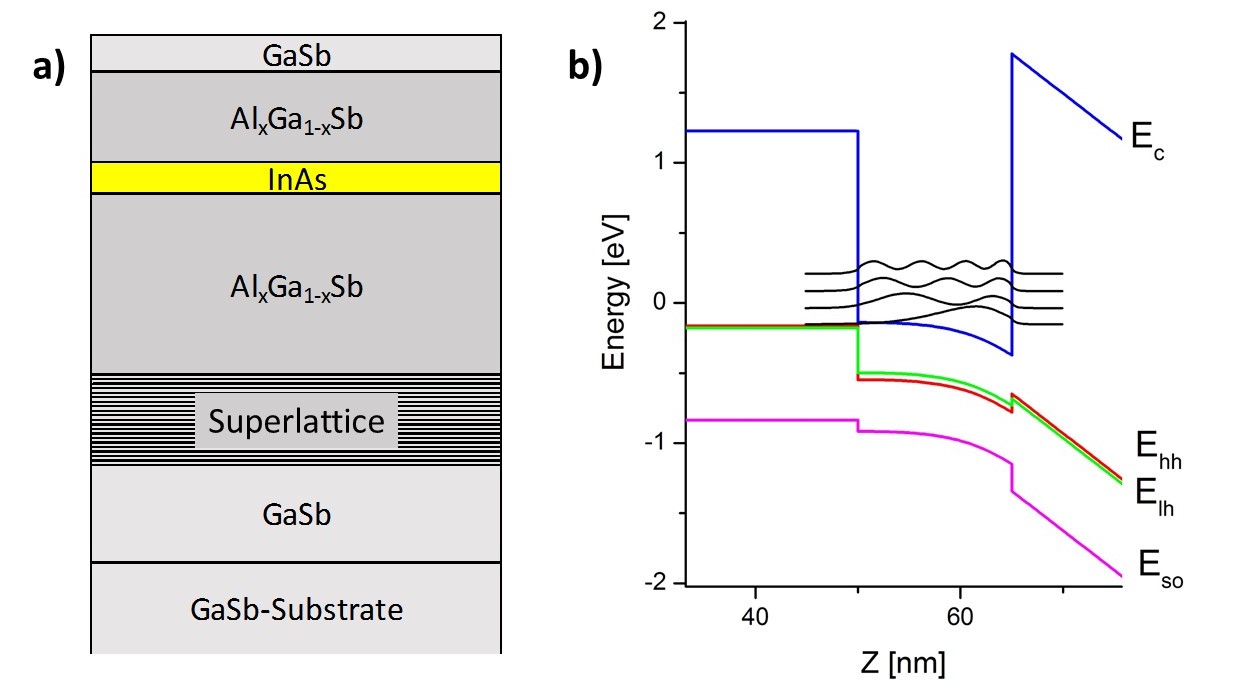}%
 \caption{(a) Illustration of the grown structures (not to scale). (b) 8$\times$8 k$\cdot$p simulation of an InAs QW with an Al$_{0.33}$Ga$_{0.67}$Sb lower barrier (left) and an AlSb upper barrier (right). E$_\textup{c}$ denotes the conduction band, E$_\textup{hh}$ the heavy hole band, E$_\textup{lh}$ the light hole band, and E$_\textup{so}$ is the split-off hole band.\label{struktur}}
 \end{figure}

After the oxide was completely desorbed and typical GaSb RHEED patterns were clearly visible, the growth on all samples was initiated with a 600 nm thick GaSb layer grown at 530 $^\circ$C. This layer was followed by a 10 period Al$_{0.33}$Ga$_{0.67}$Sb/GaSb superlattice. Towards the end of the ensuing 200 nm lower barrier, the substrate temperature was lowered to 425 $^\circ$C and the InAs quantum well was grown using the shutter sequence proposed by Tuttle \textit{et al.}\cite{Tuttle1990}. Before and after the InAs layer, when the group V element changes from Sb to As and back again, we introduced a 30 second growth pause to reduce the group V element intermixing and the consequential alloy scattering in the samples. During the InAs layer growth the temperature was increased again towards 480~$^\circ$C and the following 20 nm upper barrier and 5 nm GaSb cap layer were again grown at 530 $^\circ$C.

 \begin{figure}
 \includegraphics[width=0.495\textwidth]{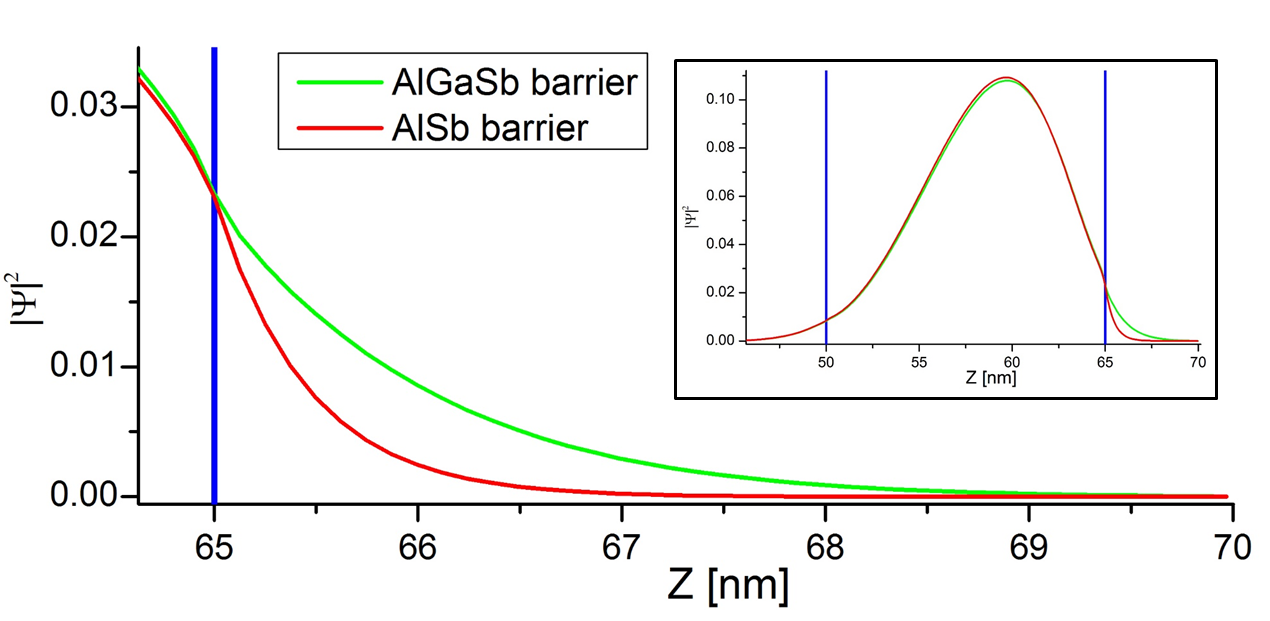}%
 \caption{8$\times$8 k$\cdot$p simulation of  a 15 nm wide InAs quantum well with either AlSb or Al$_{0.33}$Ga$_{0.67}$Sb upper barriers. The blue line indicates the onset of the barrier; the green and red lines are the squared wave functions of the lowest electron levels. The higher confinement potential of the AlSb barriers leads to a reduced penetration into the barrier compared with Al$_{0.33}$Ga$_{0.67}$Sb barriers. Inset: The simulated squared wave functions for the whole quantum well range.\label{barrier_penetration}}
 \end{figure}

We first grew a series of samples with a quantum well width of 15 nm, AlSb upper barrier, and varying lower barrier composition listed in Table \ref{15nmqws}. The first three samples demonstrate the reproducibility of the achieved electron densities and mobilities. Decreasing the aluminium content in the Al$_{x}$Ga$_{1-x}$Sb buffer and barriers reduces the lattice mismatch to the GaSb substrate and to the InAs in the quantum well. But this also causes the valence band energy of the barriers to rise relative to the conduction band energy of the quantum well. Reducing the aluminium fraction roughly below $x = 0.3$ triggers a transition from a semiconductor to a semimetal, where the electrons can transfer from the Al$_{x}$Ga$_{1-x}$Sb valence band to the InAs conduction band, inducing holes in the Al$_{x}$Ga$_{1-x}$Sb layer\cite{Lo1994}. This is illustrated in a self-consistent 8$\times$8 k$\cdot$p simulation\cite{nextnano} in Fig. \ref{struktur} and the consequences emerge in the samples D and E, where the mobilities drop with lower aluminium content. Hence, we chose an aluminium fraction of $x = 0.33$ for the buffer and the lower barriers.

Also samples with Al$_{0.33}$Ga$_{0.67}$Sb upper barriers were tested, shown in Table \ref{upperbarrier}. The intrinsic electron density with this upper barrier tends to decrease. We therefore reduced the upper barrier width to compensate that effect and keep the density in the same region as the other samples. Despite the slightly higher carrier concentrations, these wells yield lower mobilities than their counterparts with equal well width. We attribute this to the lower confinement energy resulting in a larger fraction of the electron wave function being inside the upper barrier and thus increasing alloy scattering, as illustrated in an 8$\times$8 k$\cdot$p simulation shown in Fig. \ref{barrier_penetration}. The closer surface is also likely to increase remote impurity scattering.  These effects seem to outweigh the more similar lattice constants of the quantum well and the upper barrier, which is apparently of less significance than the closer matched lattice constants between the quantum well and the lower barrier. This result is also noteworthy in the context of recent experiments on high-mobility GaAs quantum wells with either AlGaAs or AlAs barriers\cite{Kamburov2016}. There, a mobility reduction was observed for GaAs/AlAs interfaces.

 \begin{table}
 \caption{Characteristics of the 15 nm wide InAs quantum wells with 20 nm wide AlSb upper barriers. Density ($n_e$) and mobility  ($\mu$) measurements were carried out on square samples using the van der Pauw technique at 1.3 K.\label{15nmqws}}
 \begin{ruledtabular}
 \begin{tabular}{c | c | c | c }
		Sample & lower barrier & $n_e$ [$10^{11}$cm$^{-2}$]  & $\mu$ [$10^5$cm$^2$/Vs] \\
		\hline
		A & Al$_{0.33}$Ga$_{0.67}$Sb & 7.536 & 8.613 \\
		B & Al$_{0.33}$Ga$_{0.67}$Sb & 7.724 & 8.207 \\
		C & Al$_{0.33}$Ga$_{0.67}$Sb & 7.458 & 8.229 \\
		D & Al$_{0.15}$Ga$_{0.85}$Sb & 7.747 & 7.088 \\
		E & GaSb & 9.609 & 1.523 \\
 \end{tabular}
 \end{ruledtabular}
 \end{table}

 \begin{table}
 \caption{Series of InAs quantum wells with Al$_{0.33}$Ga$_{0.67}$Sb lower and upper barriers. Density ($n_e$) and mobility ($\mu$) measurements were carried out on square samples using the van der Pauw technique at 1.3 K. $W_{QW}$ denotes the quantum well width and $W_{ub}$ is the width of the upper barrier. \label{upperbarrier}}
 \begin{ruledtabular}
 \begin{tabular}{c | c | c | c | c}
		Sample & $W_{QW}$ [nm] & $W_{ub}$ [nm] & $n_e$ [$10^{11}$cm$^{-2}$]  & $\mu$ [$10^5$cm$^2$/Vs] \\
		\hline
		F & 12 & 10 & 8.200 & 2.432 \\
		G & 21 & 10 & 8.821 & 7.282 \\
		H & 15 & 50 & 3.878 & 3.724 \\
 \end{tabular}
 \end{ruledtabular}
 \end{table}

\begin{figure}
 \includegraphics[width=0.48\textwidth]{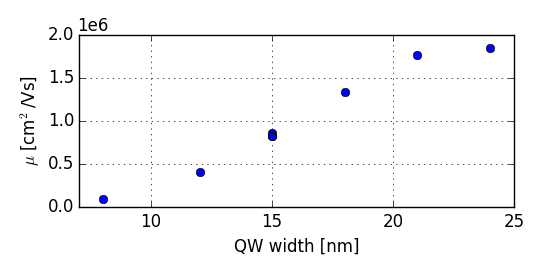}
 \caption{Plot of the electron mobilities of the samples from Table \ref{varwidthsqws} versus their quantum well width. Measurements were carried out on ungated van der Pauw samples at 1.3 K.\label{vdp_comparison}}
 \end{figure}

The samples with the lower Al$_{0.33}$Ga$_{0.67}$Sb barrier and 20 nm wide upper AlSb barrier yield the highest intrinsic mobilities and densities, the highest mobility being 0.861~$\times~10^{6}$~cm$^2$/Vs at a carrier density of 7.54~$\times 10^{11}$~cm$^{-2}$ measured at 1.3 K. We produced a series of similar samples only varying in quantum well width shown in Table~\ref{varwidthsqws}. As expected\cite{Sakaki1987,Bolognesi1992a,Bolognesi1992b}, the mobility decreases for narrowing well width due to increased interface roughness scattering. For wells wider than 15 nm, the mobility keeps increasing with an almost constant carrier density up to 1.847~$\times~10^{6}$~cm$^2$/Vs at a density of 7.99 $\times 10^{11}$ cm$^{-2}$ for the widest sample (Fig. \ref{vdp_comparison}). 

It was previously assumed that in InAs quantum wells (QWs) wider than 15 nm, the mobility starts to decrease again because the critical thickness of the InAs layer is exceeded and misfit dislocations start to nucleate\cite{Bolognesi1992a}, or due to the earlier onset of intersubband scattering\cite{Kroemer2004}. The intrinsic density of our samples is not high enough for electrons to populate the second subband even in the widest grown samples. When the lattice mismatch between the substrate and the crystal epitaxially grown upon it is reduced, less strain energy is accumulated and therefore the critical layer thickness at which the structure relaxes to the substrate lattice constant is extended\cite{Pohl}. Together with the improved growth on GaSb substrates\cite{Bennett1998} it seems possible to grow wider quantum wells and further reduce the mobility-limiting interface roughness scattering.

 \begin{table}
 \caption{Characteristics of the InAs quantum wells with Al$_{0.33}$Ga$_{0.67}$Sb lower and 20 nm AlSb upper barriers and variable quantum well width. Density ($n_e$) and mobility ($\mu$) measurements were carried out on square samples using the van der Pauw technique at 1.3~K. $W_{QW}$ denotes the quantum well width.\label{varwidthsqws}}
 \begin{ruledtabular}
 \begin{tabular}{c | c | c | c}
		Sample & $W_{QW}$ [nm] & $n_e$ [$10^{11}$cm$^{-2}$]  & $\mu$ [$10^5$cm$^2$/Vs] \\
		\hline
		I & 8  & 6.311 & 0.871 \\
		J & 12 & 7.342 & 4.083 \\
		A & 15 & 7.536 & 8.613 \\
		K & 18 & 8.151 & 13.44 \\
		L & 21 & 7.208 & 17.58 \\
		M & 24 & 7.994 & 18.47 \\
 \end{tabular}
 \end{ruledtabular}
 \end{table}

The increase of mobility with well width flattens from 21 nm to 24 nm. For a further increased well width the improvements are expected to be minimal, because all wider wells would effectively be InAs/AlSb single-interface quantum wells with a triangular well for the first electron level, as shown in the simulation in Fig.~\ref{struktur}~b).

\section{Boundary scattering in narrow Hall bars}

Hall bars were fabricated out of the two widest quantum well samples L and M using dry or wet etching as in Pal \textit{et al.}\cite{Pal2015}. Those were passivated with a 200 nm thick Si$_3$N$_4$ dielectric and covered with a Ti/Au top gate\cite{Mueller2015}.

In a magnetic field, electron trajectories are deflected by the Lorentz force. When the resulting curvature is sufficient and the elastic mean-free path $l_e$ of the electrons becomes comparable to or larger than the effective electronic width $W_{electronic}$ of a long Hall bar ($L > l_e$), the scattering with the edge becomes more relevant than scattering with potential fluctuations in the bulk of the sample. This boundary scattering appears in the measurements as an enhanced longitudinal resistivity at small magnetic fields\cite{Ditlefsen1966,Beenakker1989,Thornton1989}, peaking at $B_{max}$ according to

\begin{equation}
\frac{W_{electronic}}{R_c} = \frac{W_{electronic}eB_{max}}{\hbar k_F} = 0.55
\label{w_electronic_eq}
\end{equation}

Here, $R_c$ is the cyclotron radius and $k_F = \sqrt{2\pi n_e}$ is the Fermi wave vector.

\begin{figure}
 \includegraphics[width=0.495\textwidth]{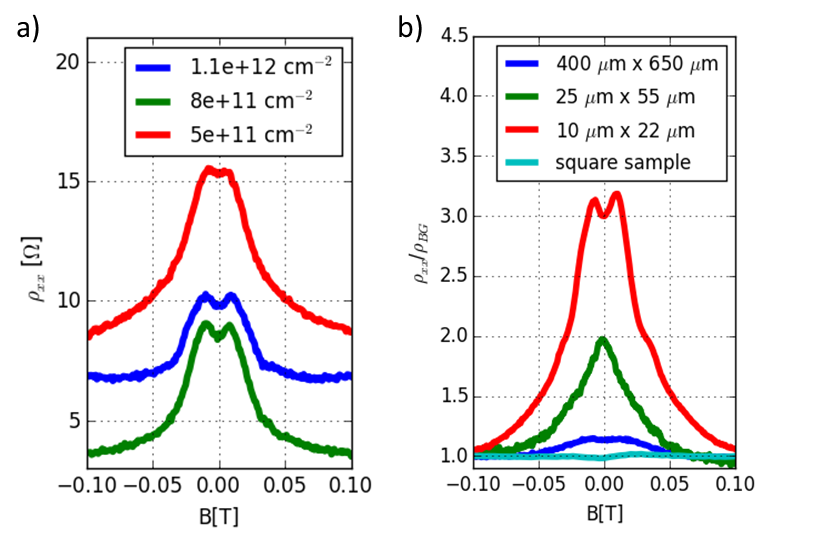}%
 \caption{(a) Peaks in longitudinal resistivity around 0 T for the 10 $\mu$m wide Hall bar device of sample M at different densities. (b) Relative enhancement of the longitudinal resistivity around 0 T for different Hall bar sizes of sample L at similar electron densities.\label{0Tpeaks}}
 \end{figure}

Measurements of the 10 $\mu$m  wide Hall bars show peaks at finite $B_{max}$ in both processed wafers [Fig.~\ref{0Tpeaks}(a)], whose position scales with $\sqrt{n_e}$ according to Eq. (\ref{w_electronic_eq}). For the 25 $\mu$m wide Hall bars the peaks are significantly smaller compared to the background resistivity and are even more reduced for the 400 $\mu$m wide Hall bars [Fig.~\ref{0Tpeaks}(b)]. In the ungated square samples the peaks have not been detected. Therefore we suspect the zero-field peaks to be an effect of the device size and most likely the boundary scattering introduced before.

 \begin{figure}
 \includegraphics[width=0.48\textwidth]{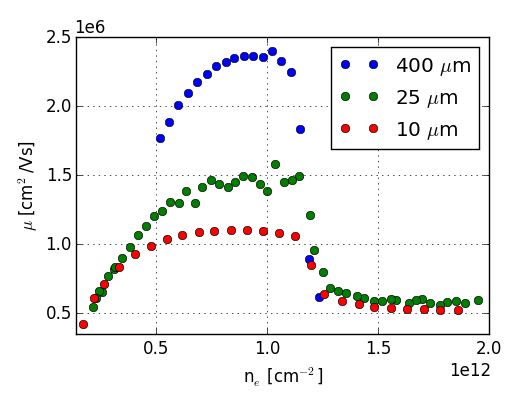}%
 \caption{Measured electron mobilities at 1.3 K of sample L with different Hall bar widths.\label{wrongmobilities}}
 \end{figure}

The enhanced zero-field longitudinal resistivity affects the calculated mobility versus density graph in a nonlinear way due to the density and mobility dependencies of the resistivity peaks (Fig. \ref{wrongmobilities}). For smaller Hall bars with presumably stronger boundary scattering the mobility curve starts to flatten at lower densities than in wider Hall bar devices. As the zero-field peak did not vanish completely in the 400~$\mu$m wide Hall bar it is likely that the flattening of the corresponding mobility curve at higher electron density is caused by residual boundary scattering also in this device.

The mean-free path at the highest mobility was $l_{e,L}~=~40~\mu$m for sample~L and $l_{e,M}~=~35~\mu$m for sample~M, calculated with\cite{Ihn} $l_e=h/\rho_{xx,0} e^2k_F$. This was measured using the zero-field longitudinal resistivities $\rho_{xx,0}$ from the widest Hall bar devices. Although these values are one order of magnitude shorter than the Hall bar width, still some residual boundary scattering seems to influence the measurements and enhances the low-field longitudinal resistance and thereby lowers the measured mobility. In the following analysis we used the measurements with the 400 $\mu$m wide and 650 $\mu$m long Hall bars where the zero-field peaks are the smallest.

The influence of boundary scattering in InAs devices has previously only been relevant for extremely narrow Hall bar devices\cite{Pal2015}. With increasing sample quality this additional scattering source also manifests itself by limiting the electron mobility in standard Hall bar sizes and has to be taken into account when fabricating devices. 

\section{Intrinsic scattering mechanisms}

\subsection{Electron mobility}

 \begin{figure*}
 \includegraphics[width=0.99\textwidth]{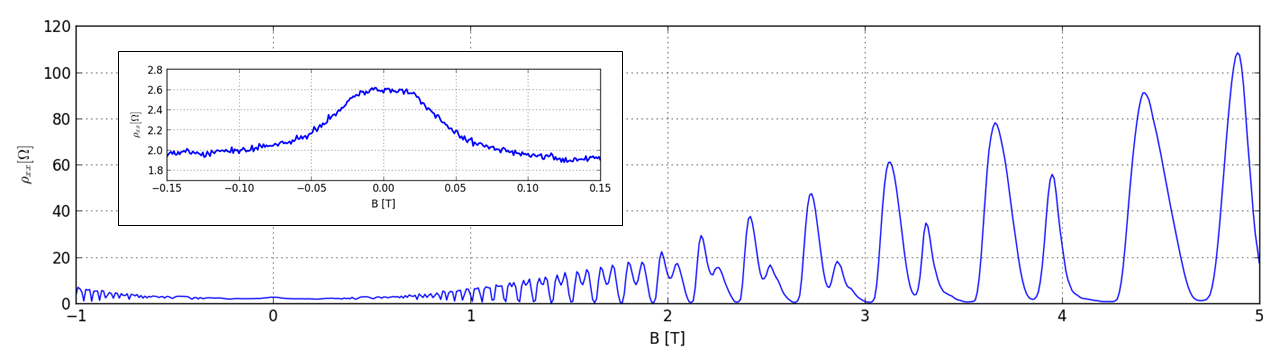}
 \caption{The longitudinal resistivity $\rho_{xx}$ of sample L versus perpendicular magnetic field $B$ at a temperature of 1.3 K and an electron density of 10.19~$\times~10^{11}$~cm$^{-2}$. Inset: Zoom-in of the same measurement around a magnetic field of 0 T. \label{sdh}}
 \end{figure*}

The carrier mobility in high-mobility electronic systems typically shows a power law dependence on the carrier density $\mu \propto n^{\alpha}$. The exponent of scaling, called the $\alpha$ parameter, is a useful tool to characterize the dominant scattering mechanisms in those systems\cite{dassarma2013}.

 \begin{figure}
 \includegraphics[width=0.49\textwidth]{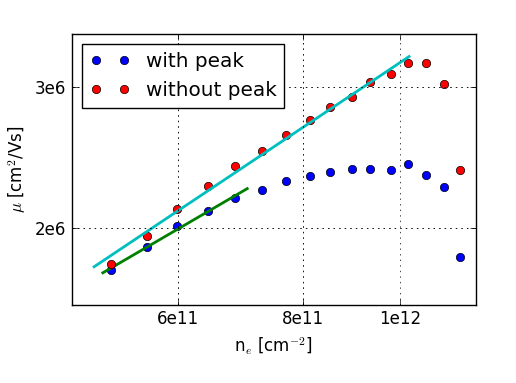}%
 \caption{The electron mobility versus electron density curve of sample L calculated before and after subtraction of the zero-field peak in logarithmic scale. The lines are the corresponding power law fits with $\alpha = 0.73$ before peak removal and $\alpha = 0.83$ after.\label{fits}}
 \end{figure}

The dependence of the mobility on electron density for sample L is shown in Fig. \ref{wrongmobilities}.  A fit to the logarithm of the low-density mobility reveals an $\alpha$ parameter of roughly 0.73 before the curve flattens. However, the accuracy of this value is limited by the small fitting range in density. This is fairly close to the theoretically predicted $\alpha$ parameter for two-dimensional systems with 3D-distributed 3D Coulomb disorder in the strong-screening limit\cite{dassarma2013} of 0.5. It also aligns nicely with the exponents of 0.7 to 0.8 reported for high-mobility GaAs/AlGaAs heterostructures\cite{Shayegan1988,Pfeiffer1989,Gold1989,Umansky1997}. In these structures such an exponent is commonly associated with dominant scattering by ionized background impurities.

At higher densities the curve flattens and reaches a maximum value of 2.402~$\times 10^{6}$~cm$^2$/Vs at a carrier density of 10.19 $\times 10^{11}$ cm$^{-2}$. This could be a sign of interface roughness scattering becoming dominant or the influence of the residual boundary scattering studied in the previous chapter. Because the boundary scattering leads to an additional longitudinal resistivity only for small magnetic fields and the remaining longitudinal resistivity is constant until the onset of Shubnikov de Haas oscillations (Fig. \ref{sdh}), it can be artificially removed in the data by calculating the mobilities using the longitudinal resistivity values at small magnetic fields beyond the onset of the peaks. When this is done using the longitudinal resistivity at a magnetic field of 0.11 T, the mobility curve follows an almost straight line until reaching a maximum value of 3.219~$\times 10^{6}$~cm$^2$/Vs and then dropping abruptly (Fig. \ref{fits}). The $\alpha$ parameter for the low to intermediate density regime in this case is 0.83, still consistent with dominant scattering by ionized background impurities. From the absence of a flattening in this case we infer that the boundary scattering is the limiting factor in this density regime in our devices.

At an electron density of approximately 11~$\times~10^{11}$~cm$^{-2}$ for sample L and 9 $\times 10^{11}$ cm$^{-2}$ for sample M a sudden drop in mobility reveals the onset of intersubband scattering, where the electrons start to fill the second subband. These density values correspond to subband separation energies of $E_{sep,L}~=~0.070$ eV and $E_{sep,M}~=~0.059$~eV using\cite{Brana2000} $E_F - E_1 = \pi\hbar^2n_e/m^*$, consistent with our 8$\times$8 k$\cdot$p simulations. The effective masses used in this calculations were $0.0375 m_0$ and $0.0365 m_0$ for the corresponding densities of samples L and M, determined as shown in the next chapter.

\subsection{Quantum scattering time}

The quantum scattering time $\tau_q$ is related to the broadening of the Landau levels and characterizes the momentum relaxation of a quasiparticle in two-dimensional transport. Together with the transport scattering time $\tau_t$ it can reveal the long-range or short-range nature of the existing scattering potentials. We measured the low-field Shubnikov–de Haas oscillations of the two highest-mobility samples L and M at different gate voltages to determine the quantum scattering time $\tau_q$ with varying electron density.

To correctly calculate the quantum scattering time, one has to take the density-dependent effective mass of InAs into account. The Shubnikov-de Haas oscillations in both samples were measured at different gate voltages over a temperature range from 1.3 K to 20 K. The effective mass was calculated via the temperature-dependence of the oscillations using the method from Bra\~na et al\cite{Brana2000}. The envelope of the oscillations in longitudinal resistivity is given by \cite{Ihn,Coleridge1989}

\begin{equation}
\frac{\Delta \rho_{xx}}{\bar{\rho}_{xx}} = \pm 2 \exp[-\pi/(\omega_c \tau_q)] \frac{\chi(T)}{\sinh[\chi(T)]},
\end{equation}

where $\chi(T) \equiv 2\pi^2k_BT/\hbar \omega_c$, $\omega_c = eB/m^*$ is the cyclotron frequency, and $\bar{\rho}_{xx}$ is the nonoscillatory background resistivity. The amplitude of the oscillations scales with temperature as

\begin{equation}
A(B,T) \propto \frac{\chi}{\sinh[\chi(T)]}.
\end{equation}

Using $\sinh(x) = \frac{1-e^{-2x}}{2e^{-x}}$, we can define

\begin{equation}
f(T;m^*) \equiv \ln \left(\frac{A}{T}\right) + \ln \left[1-\exp\left(-\frac{4\pi^2k_BT}{\hbar \omega_c} \right)\right] 
\end{equation}

and it follows that

\begin{equation}
f(T;m^*) \propto - \frac{2\pi^2 k_B T m^*}{eB}.
\end{equation}

When $f(T;m^*)$ is calculated from the measured amplitudes and plotted against the temperature $T$, the effective mass $m^*$ is incorporated in the slope as well as in $f(T;m^*)$ itself. This equation has to be solved repeatedly until both effective masses match. 

We calculated the effective masses from all peaks in the magnetic field range where zero-field spin-splitting beatings do not occur and before the onset of spin-orbit coupling induced spin splitting, which is roughly between 0.8 T and 1.8 T depending on sample and electron density. The calculated values did not show any significant magnetic field dependence in the respective ranges.

\begin{figure}
 \includegraphics[width=0.47\textwidth]{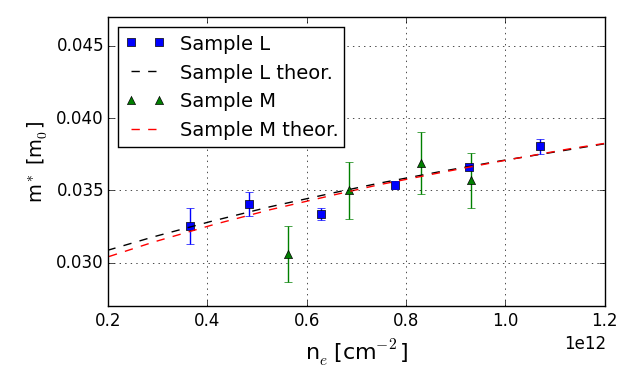}%
 \caption{The effective masses $m^*$ of samples L and M at different electron densities $n_e$. The dashed lines are the theoretical calculations for 21 nm (black) and 24 nm (red) wide InAs quantum wells as described in the text.\label{effmasses}}
 \end{figure}

The average of $m^*$ for all peaks at the corresponding electron densities is shown in Fig. \ref{effmasses}. The dashed lines are the effective masses calculated from the bulk effective mass using the formula derived by Ando\cite{Ando1982},

\begin{equation}
\frac{\Delta m^*}{m^*_b} = \sqrt{1+4\frac{(\left\langle K \right\rangle + E_F)}{E_g}}-1,
\end{equation}

 where $E_F$ is the Fermi energy, $E_g$ the band gap energy, $m^*_b$ the bulk effective mass, and $\Delta m^*$ is the enhancement of the effective mass in a quantum well over the bulk effective mass. The kinetic energy $\left\langle K \right\rangle$ can be approximated for a triangular well with $E_1/3$, where $E_1$ is the energy of the first subband calculated using 8$\times$8 k$\cdot$p simulations. The Fermi energy is given by $E_F = \pi \hbar^2 n_e / m^*$. The bulk effective mass for the theoretical calculation fitting the measurement best was $0.024 m_0$, which lies comfortably in the range of reported values\cite{Vurgaftman2001} for InAs. 

The measured masses match the model very well. Besides the higher precision of the measurements of sample L, there was no clear difference between the two well widths.  This is in accordance with the theoretical calculations, in which the masses of the two quantum wells with different width deviate slightly from each other only at low densities.

These masses were now used in the calculation of the quantum scattering time $\tau_q$. By plotting 

\begin{equation}
\ln \left(\frac{\Delta \rho_{xx}\sinh[\chi(T)]}{4\bar{\rho}_{xx}\chi(T)}\right) = -\frac{\pi m^*}{e\tau_q B} + const.
\end{equation} 

versus $1/B$ at constant temperature, one can calculate $\tau_q$ from the slope of the resulting straight lines\cite{Ihn}. Only the results of valid fits are shown\cite{Coleridge1991}, where the plots are straight lines and the intercepts at $1/B = 0$ are close to~1.

\begin{figure}
 \includegraphics[width=0.5\textwidth]{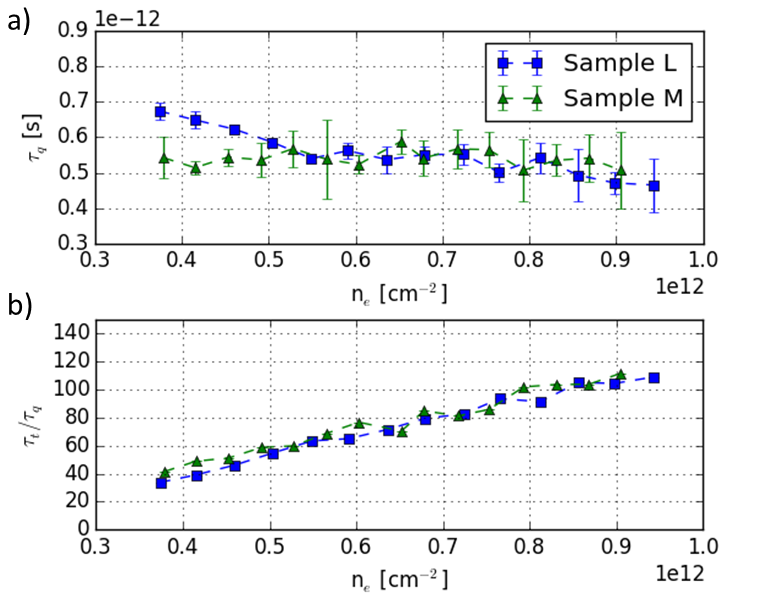}
 \caption{(a) Measured quantum scattering time $\tau_q$ for samples L and M at different electron densities $n_e$. (b) Dingle ratio of the samples L and M for varying electron density $n_e$.\label{tau_q}}
 \end{figure}

The measured quantum scattering time for both samples at 1.3 K is presented in Fig. \ref{tau_q} a). It fluctuates for both samples roughly around $5.5 \times 10^{-13}$~s. For sample M $\tau_q$ is constant over the whole density range. For sample~L it exhibits a slight apparent decrease with increasing carrier density. Such a decrease can be caused by interface roughness scattering\cite{Lo1992}. We deem this unlikely, because the apparent decrease in our sample occurs over the whole gate range instead of only at high densities where interface roughness scattering is supposed to dominate.

The Dingle ratio, defined as $\tau_t$/$\tau_q$, where $\tau_t = \mu m^*/e$ is the transport lifetime, can be taken as an indication of the nature of the the scattering potential\cite{Gold1988,Peters2016}. $\tau_t$ is weighted heavily towards large-angle scattering events and $\tau_q$ is affected by scattering events of all angles. For short-range isotropic scattering both lifetimes should be almost equally affected and a ratio close to 1 is expected. For scattering by long-range Coulomb interactions $\tau_t$ is only weakly affected and the ratio $\tau_t$/$\tau_q$ is considerably larger.

Figure \ref{tau_q}(b) shows the measured Dingle ratio for both samples. At low densities both samples have a Dingle ratio of roughly 40, which is comparable to reported Dingle ratios in high-mobility GaAs and in SiGe quantum wells\cite{Fang1988,Coleridge1991,Ismail1995}. For higher densities the Dingle ratios of both samples increase in unison, indicating even smoother scattering potentials. We explain this by the extremely high electron density shielding remote scatterers and thus smoothing out their scattering potential. The slight decrease of $\tau_q$ in sample L has no significant influence on this ratio as the result of both samples is very similar.

As a result of this analysis we suspect the dominant scattering in this density regime to stem from long-range Coulomb interactions with ionized impurities predominantly remote to the well. The boundary scattering from the previous chapter does not appear in this analysis because $\tau_q$ is measured at higher magnetic fields where the zero-field peak in longitudinal resistance is irrelevant.

\section{Landau fan}

\begin{figure}
 \includegraphics[width=0.5\textwidth]{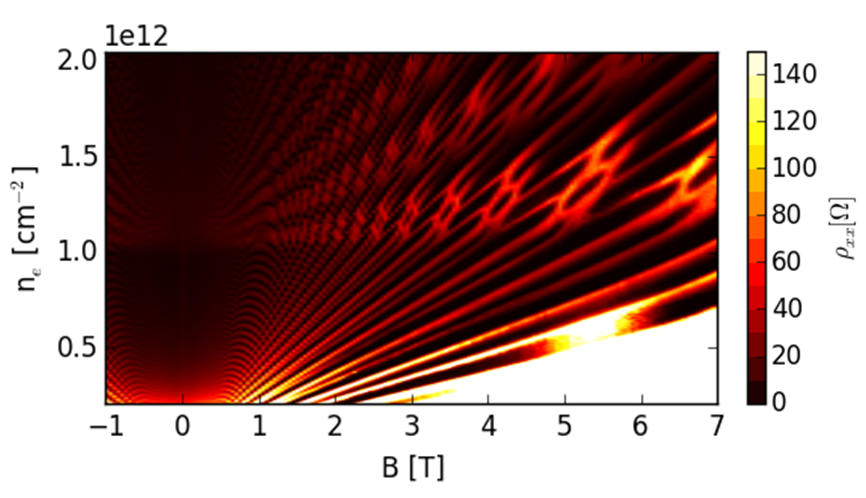}%
 \caption{Landau fan of sample M. Ringlike features appear where the Landau levels of the first subband intersect the Landau levels of the second subband.\label{bigfan}}
 \end{figure}

In the high electron density regime, where two subbands are occupied, both samples show peculiar, ringlike features in their Landau fans (Fig.~\ref{bigfan}). Similar features were also observed in GaAs/AlGaAs heterostructures\cite{Muraki2001,Zhang2005,Ellenberger2006,Zhang2006,Gusev2007} with two occupied subbands or in a tilted magnetic field. Despite the usually higher single subband mobilities of the GaAs/AlGaAs system, these samples only reached a mobility of about 0.6~$\times 10^{6}$~cm$^2$/Vs or less in the two-subband regime due to strong intersubband scattering and heavy doping. The studied InAs samples exhibit a comparable mobility in their second subband regime and the ringlike features are clearly visible at measurement temperatures of~1.3~K.

These peculiar structures can be reasonably reproduced using a single-particle model introduced by Ellenberger et al. for GaAs/AlGaAs heterostructures\cite{Ellenberger2006}. The evolution of the Landau level energy in a magnetic field is given by

\begin{eqnarray}
E_{1,n,s} &= &\left(n+\frac{1}{2}\right)\hbar\omega_c + s\frac{1}{2}g^*\mu_BB \\
E_{2,n,s} &= &\Delta E + \left(n+\frac{1}{2}\right)\hbar\omega_c + s\frac{1}{2}g^*\mu_BB
\end{eqnarray}

for the respective subbands 1 and 2 with the Landau-level quantum number $n$ and the two spin directions $s = \pm 1$. The crucial step in this model is the highly nonlinear mapping from the energy--magnetic field plane to the density--magnetic field plane. This is achieved by assuming Gaussian-broadened Landau levels which can be maximally occupied by $eB/h$ electrons per area. After this mapping, the ringlike structures appear in the simulation.

The g-factor of $g_1 = 12.2$ for the simulation was determined from the crossing points in the Landau fan as in Ellenberger et al. The effective mass of the first electron subband $m_{e1}^* = 0.039 m_0$ at this density was determined using the theoretical curve calculated along the measurements presented earlier (Fig. \ref{effmasses}). The effective mass of the second subband in this density range was estimated using 8$\times$8 k$\cdot$p simulations as $m_{e2}^* = 0.038 m_0$. The measured quantum scattering time $\tau_q = 5.5 \times 10^{-13}$~s yields a level broadening $\Gamma$ of 600 $\mu$eV, using $\Gamma = \hbar/2\tau_q$. 

The energy separation of the first and second electron subband varies with density because the top gate influences the shape of the quantum well. This was implemented using a linear fit of the energy separations at various densities determined from crossing points in the Landau fan. The obtained energy separations are in close agreement to self-consistent 8$\times$8 k$\cdot$p simulations and with  the values calculated from the onset of the second subband in the mobility curve.

\begin{figure}
 \includegraphics[width=0.49\textwidth]{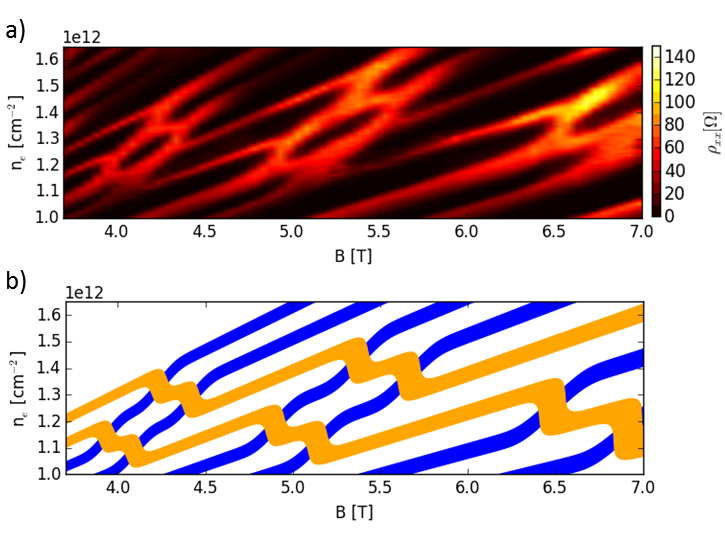}
 \caption{(a) A region in the Landau fan of sample M with two occupied electron subbands. (b) The simulation of the same region as described in the text. The blue lines are Landau levels of the first electron subband, and the orange lines are the spin-split first Landau level of the second electron subband.\label{landausimcomparison}}
 \end{figure}

Figure \ref{landausimcomparison} shows the simulated crossing of the three spin-split Landau levels $n = 3,4,5$ from the first subband with the $n = 0$ Landau level from the second subband and the corresponding region in the measurement. This single-particle picture matches the experiment remarkably well with input parameters only from measurement and 8$\times$8 k$\cdot$p~simulations.

\section{Conclusion}

We have grown a range of undoped InAs quantum wells with different buffer and barrier structures and quantum well widths using molecular beam epitaxy. We found that lower Al$_{x}$Ga$_{1-x}$Sb buffers and barriers with $x = 0.33$ yield the highest intrinsic electron mobilities. Extending the quantum well width beyond the widely used 15~nm towards 24 nm yields another dramatic improvement in mobility. 8$\times$8 k$\cdot$p simulations show that these widest structures are quasi-single-interface InAs/AlSb quantum wells. The highest ungated mobility achieved was 1.847~$\times 10^{6}$~cm$^2$/Vs at a density of 7.99 $\times 10^{11}$ cm$^{-2}$.

By manufacturing Hall bars with top gates we were able to tune the electron density and increase the mobility to a maximum of 2.402~$\times 10^{6}$~cm$^2$/Vs at a density of 10.19~$\times~10^{11}$~cm$^{-2}$. We found the size of these Hall bars to heavily influence the measured mobilities. Since the elastic mean-free path $l_e$ of the electrons at mobilities this high easily reaches the width of standard Hall bars, scattering at the device boundaries enhances the low magnetic field longitudinal resistivity considerably.

The density-dependent effective mass of the highest-mobility samples was measured and it agrees with theoretical calculations. The measured quantum scattering time $\tau_q$ and its relation to the transport scattering time $\tau_t$ of these samples indicate that their most important sources of scattering for a nonzero magnetic field and before the population of the second subband are likely to be ionized impurities predominantly remote to the well. This is supported by the scaling exponent $\alpha$ relating the carrier mobility to the carrier density. For lower densities it has a value of $\alpha = 0.73$ consistent with dominant scattering by background impurities as in high-mobility GaAs/AlGaAs structures. For higher densities scattering at the device boundaries becomes the limiting factor and at even higher densities the onset of intersubband scattering causes a sudden drop in mobility.

We also encountered ringlike features in the Landau fans of our two highest-mobility samples. These features arise from the crossing of spin-split Landau levels of two different subbands and were modeled using a single-particle picture.

\begin{acknowledgments}
The authors acknowledge financial support by the Swiss National Science Foundation (SNF) through the NCCR 'QSIT – Quantum Science and Technology' (National Centre of Competence in Research).
\end{acknowledgments}

\bibliography{InAsPaper}

\end{document}